\documentstyle[aps,preprint]{revtex}

\begin{document}
\draft
\title
{\bf Ginzburg-Landau Expansion and the Slope of the Upper Critical Field
in Disordered Superconductors with Anisotropic Pairing}
\author{A.I.Posazhennikova,\ M.V.Sadovskii}
\address
{Institute for Electrophysics,\\ Russian Academy of Sciences,\ 
Ural Branch,\\ Ekaterinburg,\ 620049, Russia\\
E-mail:\ sadovski@ief.intec.ru} 
\maketitle

\begin{center}
{\sl Submitted to JETP Letters, December 1995}
\end{center}

\begin{abstract}
It is demonstrated that the slope of the upper critical field
$|dH_{c2}/dT|_{T_{c}}$ in superconductors with  $d$-wave pairing drops rather
fast with concentration of normal impurities, while in superconductors with
anisotropic $s$-wave pairing $|dH_{c2}/dT|_{T_{c}}$ grows, and in the limit of 
strong disorder is described by the known dependences of the theory of
``dirty'' superconductors. This allows to use the measurements of $H_{c2}$ 
in disordered superconductors to discriminate between these different types
of pairing in high-temperature and heavy-fermion superconductors.
\end{abstract}
\pacs{PACS numbers:  74.20.Fg, 74.20.De}

\newpage
\narrowtext

The main problem of the present day physics of high-temperature superconductors
is the determination the nature and type of the Cooper pairing. A number of
experiments and theoretical models\cite{DP} suggest the realization in these
systems of anisotropic pairing with $d_{x^2-y^2}$-symmetry with appropriate
zeroes of the gap function at the Fermi surface. At the same time most of these
experiments also agree with the so called anisotropic $s$-wave pairing, which
follows from some theoretical models\cite{CS,LM}. In this latest case there
again appear the zeroes (with no change of sign) or rather deep minima
of the gap function in the same directions in the Brillouin zone as in the
case of  $d$-wave pairing.

Recently it was shown\cite{BH,FN} that controlled disordering (introduction of
normal impurities) can be an effective method of experimental discrimination
between different types of anisotropic pairing. Disordering leads to 
different behavior of the density of states in superconducting state: $d$-wave
pairing superconductor remains gapless, while in an anisotropic $s$-wave
superconductor with zeroes of the gap function, small disordering leads to the
opening of the finite gap on the Fermi surface.

Gap measurements, especially for the different directions in the Brillouin zone,
are difficult enough to perform. The aim of the present paper is to demonstrate
that much simpler, in principle, measurements of the upper critical field
$H_{c2}$ at different degrees of disorder can also provide an effective method
to discern $d$-wave pairing from anisotropic $s$-wave. Surely, the problem
under discussion is of interest also for heavy-fermion superconductors.

Following Refs.\cite{BH,FN},\ we analyze two-dimensional electronic system
with isotropic Fermi surface and separable pairing potential of the form:
\begin{equation}
V(\phi ,\phi ')=-V\eta(\phi)\eta(\phi') \label{1}
\end{equation}
where $\phi$-is a polar angle, determining the electronic momentum direction in
the plane, and $\eta(\phi)$ is given by the following model dependence:
\begin{equation}
\eta(\phi)=\left\{
\begin{array}{ll}
cos\left(2\phi\right) & ($d$-\mbox{ wave }) \\
|cos\left(2\phi\right)| & (\mbox{ anisotropic } $s$-\mbox{ wave })
\end{array}
\right.
\label{2}
\end{equation}
The pairing constant $V$ is as usual different from zero in some region of
the width of $2\omega_{c}$ around the fermi level ($\omega_{c}$-is some
characteristic frequency of the quanta, responsible for the pairing interaction
). In this case the superconducting gap (order parameter) takes the form:
$\Delta(\phi)=\Delta\eta(\phi)$, and positions of its zeroes for $s$ and $d$
cases just coincide.

BCS equations for the impure superconductor are derived in a standard way
\cite{AG}. Linearized gap equation, determining the transition temperature
$T_{c}$ takes the form3:
\begin{equation}
\Delta(\phi)=-N(0)T_{c}\sum_{\omega_{n}}\int_{-\infty}^{\infty}d{\xi}
\int_{0}^{2\pi}\frac{d{\phi'}}{2\pi}V(\phi, \phi')\frac{\tilde\Delta(\phi')}
{\tilde\omega_n^2+ \xi^2} \label{3}
\end{equation}
where
\begin{equation}
\tilde\Delta(\phi)=\left\{
\begin{array}{ll}
\Delta\eta(\phi) &($d$\mbox{ - wave }) \\
\Delta(\eta(\phi)+2\gamma/\pi|\omega_{n}|) &(\mbox{ anisotropic }
$s$\mbox{ - wave })
\end{array}
\right.
\label{2a}
\end{equation}
$\tilde\omega_{n}=\omega_{n}+ \gamma sign(\omega)$, 
$\gamma = \pi\rho V_{0}^2 N(0)$ - is the usual electron damping due to
impurity scattering, $V_{0}$ - impurity potential and 
$\rho$ - impurity concentration, $N(0)$ - is normal density of states at the
Fermi level and $\xi$ - is electronic energy with respect to the Fermi level,
$\omega_{n}= (2n+1) \pi T_{c}$.

After the traditional analysis $T_{c}$-equation reduces to\cite{BH,FN}:
\begin{equation}
ln\biggl(\frac{T_{c0}}{T_c}\biggr)=
\alpha\biggl[\Psi\biggl(\frac{1}{2}+\frac{\gamma}{2\pi T_{c}}\biggr)-
\Psi\biggl(\frac{1}{2}\biggr)\biggr] \label{5}
\end{equation}
where $\alpha= 1$ for the case of $d$-pairing and $\alpha= (1-8/\pi^2)$ for
anisotropic $s$-wave pairing, $T_{c0}$-is the transition temperature in the
absence of impurities, $\Psi(x)$- is the usual digamma function. The 
appropriate dependences of $T_{c}(\gamma/T_{c0})$ are shown in Fig.1.
In case of $d$-pairing $T_{c}$ is completely suppressed for
$\gamma=\gamma_{c}\approx 0.88T_{c0}$. In anisotropic $s$-case the dependence
of $T_{c}$ on $\gamma$ is much weaker, for $\gamma\gg T_{c0}$ we obtain
$T_{c}\sim T_{c0}[1-\alpha ln(\gamma/\pi T_{c0})]$.

Ginzburg-Landau expansion for the free-energy density of a superconducting
state up to terms quadratic over $\Delta_{q}$ can be written as:
\begin{equation}
F_{s}-F_{n}=A|\Delta_{q}|^2+q^2 C|\Delta_{q}|^2 \label{6}
\end{equation}
and is determined by the usual loop-expansion for the free-energy of an 
electron in the field of random fluctuations of the order-parameter with some
small wave vector $q$, shown in Fig2. Diagramms (c) and (d) are to be
subtracted, so that the coefficient $A$ becomes zero for $T=T_{c}$.
All calculations are standard and we only note that for the case of 
$d$-pairing the contribution of diagramms (b) and (d) actually vanishes up to
terms of the order of $q^4$. Finally, the coefficients of Ginzburg-Landau
expansion can be written in the following form:
\begin{equation}
A=A_{0}K_{A};\qquad   C=C_{0}K_{C}
\end{equation}
where $A_{0}$ and $C_{0}$ are the usual expressions for the case of isotropic
$s$-wave pairing\cite{PG}:
\begin{equation}
A_{0}=N(0)\frac{T-T_{c}}{T_{c}};\qquad
C_{0}=N(0)\frac{7\zeta(3)}{48\pi^{2}}\frac{v_{F}}{T_c^2}   \label{7}
\end{equation}
where $v_{F}$-is electron velocity at the Fermi surface, and all peculiarities
of models udder consideration are actually contained in dimensionless
coefficients $K_{A}$ and $K_{C}$. In the absence of impurities for both
models we obtain: $K_A^0=1/2$, $K_C^0=3/4$. For the impure system we get:

(A) $d$-wave pairing:
\begin{equation}
K_{A}= \frac{1}{8T_{c}}\int_{-\omega_{c}}^{\omega_{c}}\frac{d{\xi}}{\xi}
\int_{-\infty}^{\infty}\frac{d{\omega}}{\pi}
\frac{\omega+\xi}{ch^2\left(\frac{\omega+\xi}{2T_{c}}\right)}
\frac{\gamma}{\omega^{2}+\gamma^{2}}  \label{8}
\end{equation}
\begin{equation}
K_{C}=-\frac{3}{56\zeta(3)} \Psi''\biggl(\frac{1}{2}+\frac{\gamma}
{2\pi T_{c}}\biggr)
\label{9}
\end{equation}

(B) anisotropic $s$-wave pairing:
\begin{equation}
K_{A}=\frac{\gamma}{\pi T_{c}}\Biggl\{\frac{1}{8}\int_{-\omega_{c}}^{\omega_{c}}
\frac{d{\xi}}{\xi}\int_{-\infty}^{\infty}d{\omega}\frac{\omega+\xi}
{ch^2\left(\frac{\omega+\xi}{2T_{c}}\right)(\omega^2+\gamma^2)}+
\frac{\gamma}{\pi}\int_{-\infty}^{\infty}d{\omega}\frac{1}
{ch^2\left(\frac{\omega}{2T_{c}}\right)(\omega^2+\gamma^2)}\Biggr\} \label{10}
\end{equation}
\begin{equation}
K_{C}=-\frac{3\alpha}{56\zeta(3)}\Psi''\biggl(\frac{1}{2}+
\frac{\gamma}{2\pi T_{c}}\biggr)
+\frac{24}{7\zeta(3)}\frac{T_c^2}{\alpha\gamma^2}
ln\biggl(\frac{T_{c}}{T_{c0}}\biggr)+
\frac{6\pi}{7\zeta(3)}\frac{T_{c}}{\gamma}  \label{11}
\end{equation}
The appropriate dependences of dimensionless coefficients on disorder
parameter $\gamma/T_{c0}$ are shown in Figs.3,4.

Close to $T_{c}$ the upper critical field $H_{c2}$ is determined from (\ref{8}):
\begin{equation}
H_{c2}= -\frac{\phi_{0}}{2\pi}\frac{A}{C} \label{12}
\end{equation}
where $\phi_{0}=c\pi/e$ --- is magnetic flux quantum. Then the slope of the
upper critical field close to $T_{c}$ is:
\begin{equation}
\left|\frac{dH_{c2}}{dT}\right|_{T_c}=\frac{24\pi\phi_{0}}{7\zeta(3)v_F^2}T_{c}
\frac{K_A}{K_C} \label{13}
\end{equation}
Dependence of $|dH_{c2}/dT|_{T_c}$ on $\gamma/T_{c0}$ for both models is
shown in Fig.5. We can see that for the case of $d$-wave pairing
the slope of $H_{c2}$ drops to zero on the scale of $\gamma\sim T_{c0}$. 
For the case of anisotropic $s$-wave pairing, on the contrary, the slope
grows with disorder and after some transition region of $\gamma\sim T_{c0}$ it
crosses over to the usual linear dependence
$|dH_{c2}/dT|_{T_c}\sim\gamma$, which is characteristic of the usual theory of
``dirty'' superconductors with isotropic $s$-wave pairing\cite{G}. 
In our opinion this sharp difference can be used a simple enough criterion of
experimental discrimination of $d$-wave superconductors from anisotropic
$s$-wave case. Unfortunately, in case of high-$T_{c}$ oxides the situation is
complicated by the known nonlinearity of temperature dependence of
$H_{c2}$, which is observed in rather wide region close to $T_{c}$. 
At present the nature of this nonlinearity is unclear and it is probable that
it may be due to some inhomogeneity of the samples, though its more 
fundamental nature can not be excluded.

This work was performed under the project No.93-001 of the State Program of
Research on HTSC. It was also partly supported by the grant 93-02-2066 of
the Russian Foundation of Fundamental Research, as well as by the grant of
International (Soros) Science Foundation RGL300. The authors are grateful for
this support of their research.

\newpage

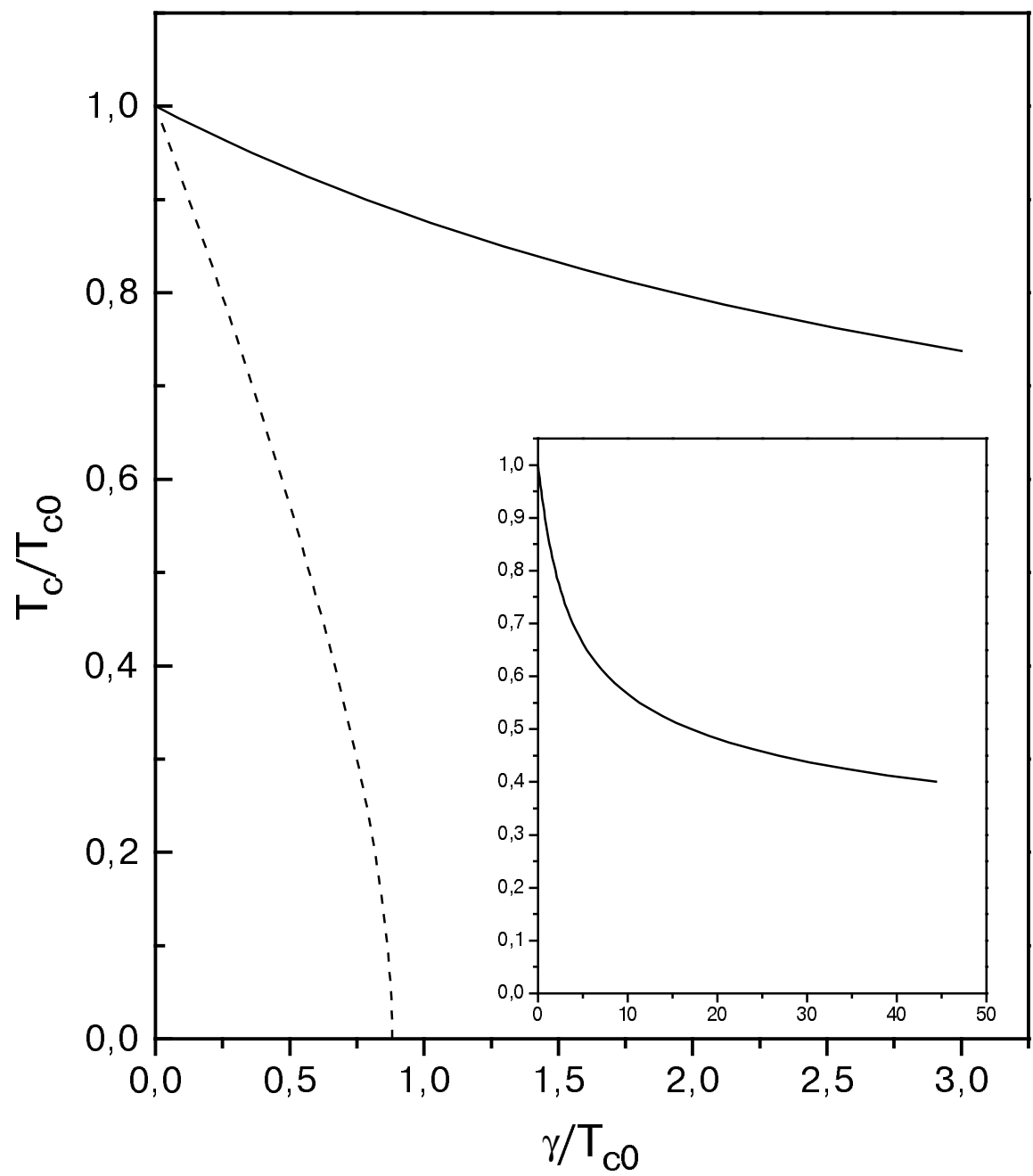
Fig.1. Dependence of transition temperature $T_{c}$ on disorder parameter
$\gamma/T_{c0}$. Dashed line  - dependence for the case of
$d$-wave pairing, full line - the case of anisotropic $s$-wave pairing. 
At the insert - the same curve for $s$-pairing for wider interval of the
parameter $\gamma/T_{c0}$.

Fig.2. Diagrammatic representation of Ginzburg-Landau expansion. Electronic
lines are ``dressed'' by impurity scattering. $\Gamma$-is the impurity vertex
calculated in ``ladder'' approximation. Diagramms (c) and (d) are calculated
with $q=0$ and $T=T_{c}$.

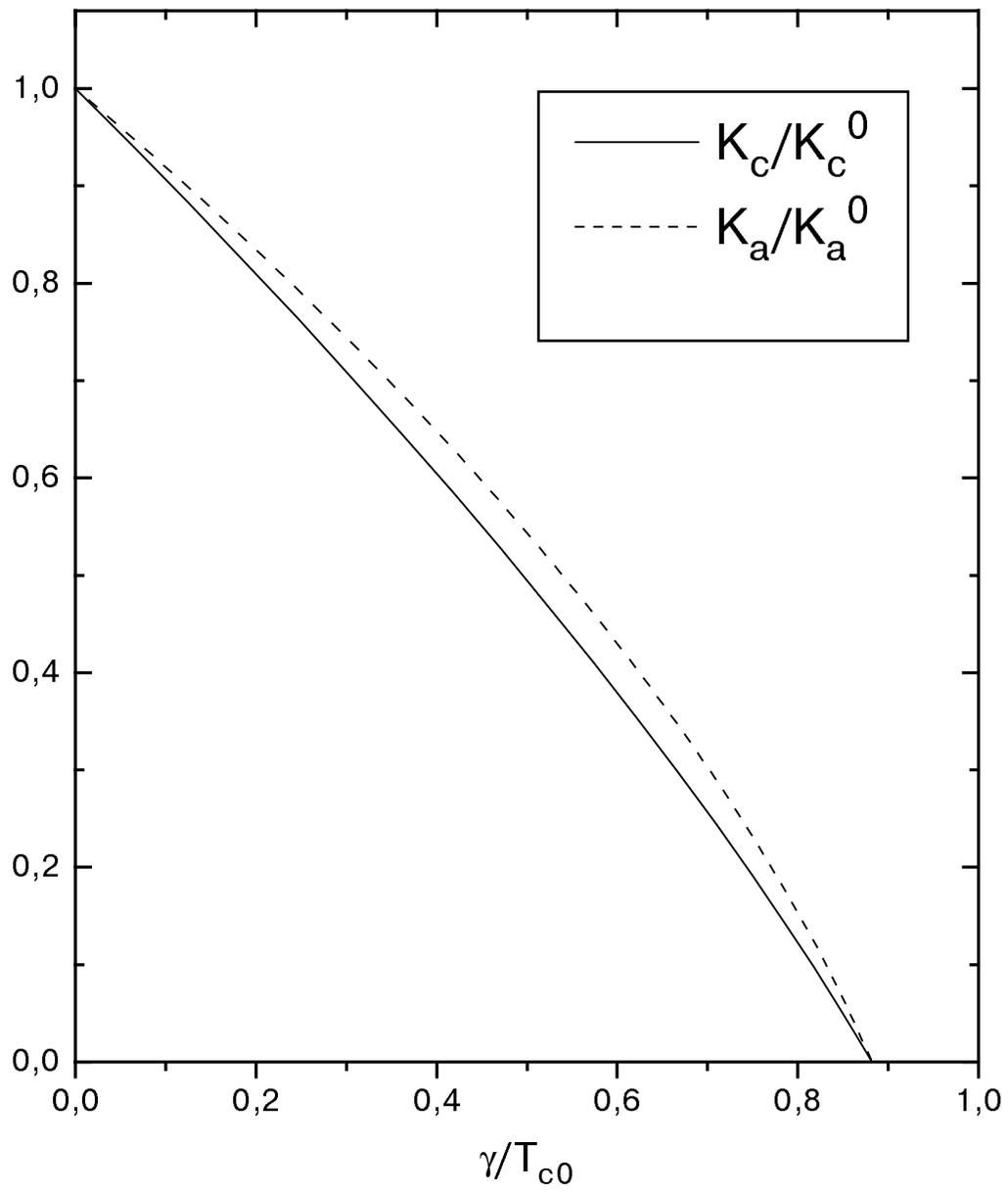
Fig.3. Dependence of dimensionless coefficients of Ginzburg-Landau expansion
on disorder parameter $\gamma/T_{c0}$. The case of $d$-wave pairing.

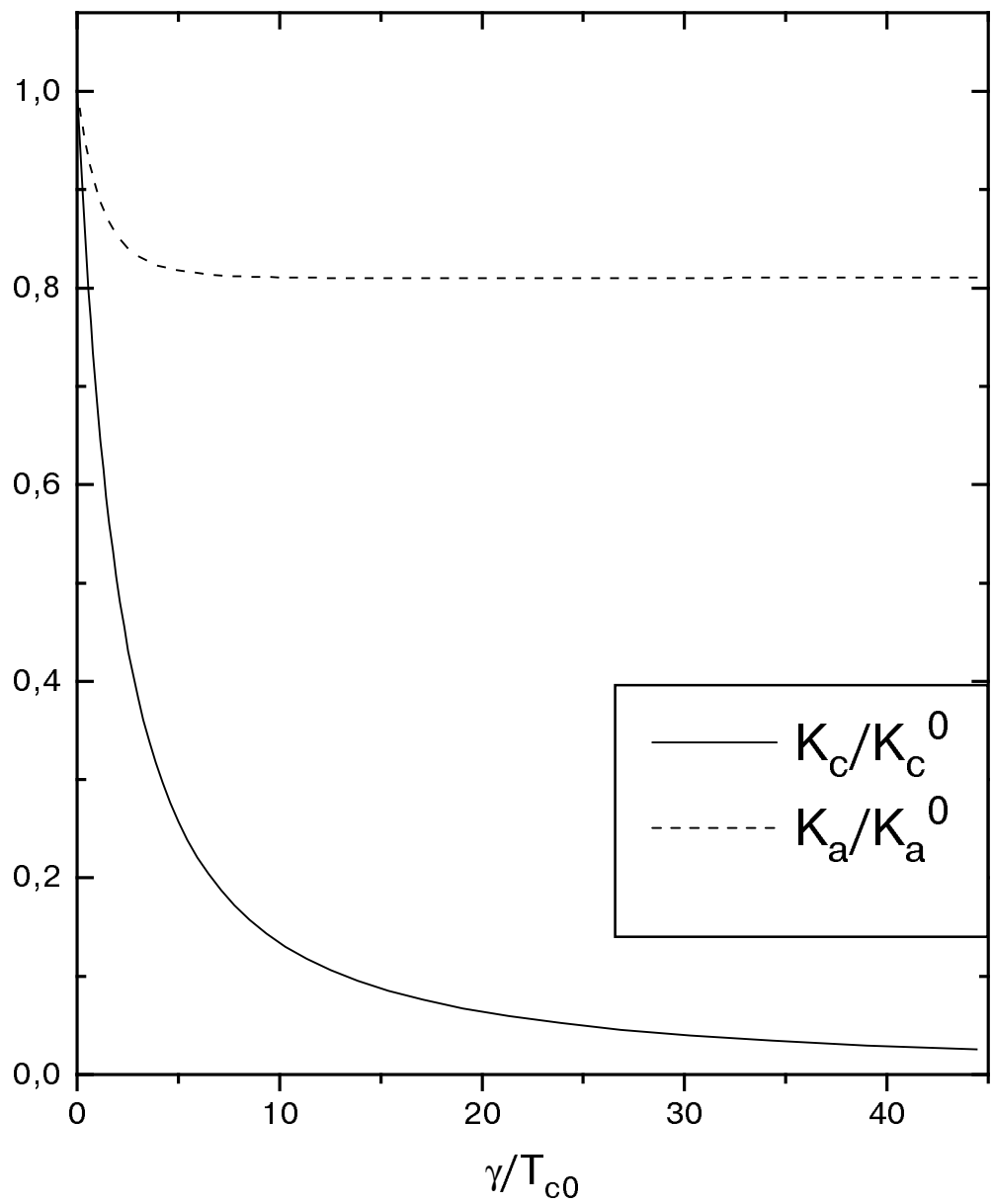
Fig.4. Dependence of dimensionless coefficients of Ginzburg-Landau expansion
on disorder parameter $\gamma/T_{c0}$. The case of anisotropic $s$-wave
pairing.

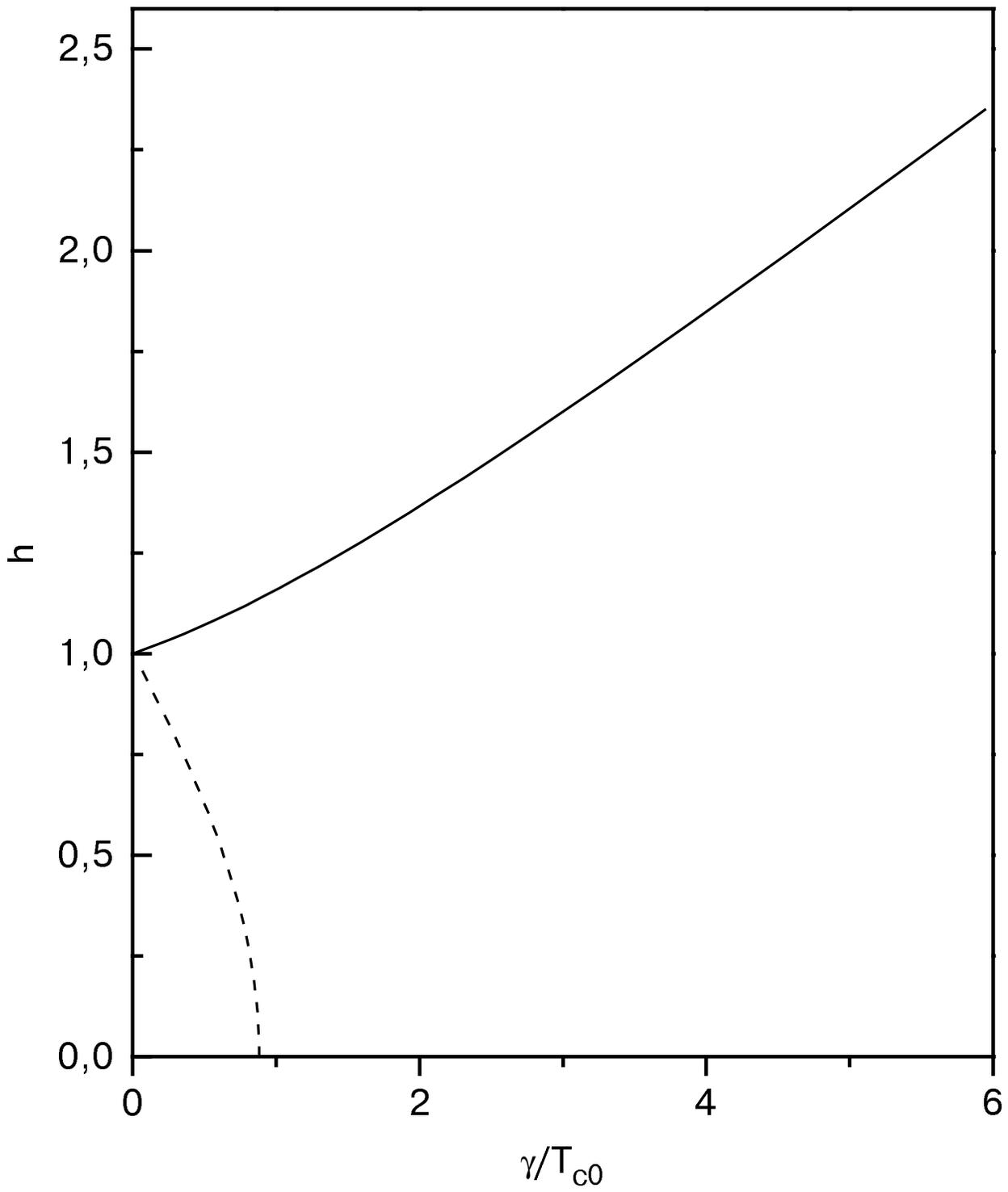
Fig.5. Dependence of normalized slope of the upper critical field
$h=\left|\frac{dH_{c2}}{dT}\right|_{T_c}/
\left|\frac{dH_{c2}}{dT}\right|_{T_{c0}}$ on disorder parameter
$\gamma/T_{c0}$. Dashed line - the case of $d$-wave pairing, 
full line - the case of anisotropic $s$-wave pairing.

\end{document}